\documentclass{article}

\usepackage{graphicx}
\usepackage{wrapfig}
\usepackage{authblk}

\begin{document}
\title{Search for Pentaquark $\Theta^{+}$ in Hadronic Reaction at J-PARC
%\thanks{Presented at ...}%
}

\author{Megumi Naruki for the J-PARC E19 Collaboration}
\affil{Kyoto University}
%\address{Kyoto University}

\maketitle

\begin{abstract}
  A variety of nuclear and hadron physics experiments are being performed
  using meson beams at the J-PARC Hadron Facility.
  As the first experiment at the facility,
  the pentaquark $\Theta^{+}$ was searched for in the $\pi^{-}p \rightarrow
  K^{-}X$ reaction with a missing-mass resolution of 2~MeV (FWHM).
  The number of accumulated beam pions are $7.8\times 10^{10}$ and $8.1\times 10^{10}$ for different beam 
  momenta of 1.92 and 2.01~GeV/$c$, respectively.
  No significant structure was observed in the missing mass spectra.
  Upper limits of the production cross section are obtained to be $0.28\mu$b/sr
  in the laboratory frame
  at $90\%$ confidence level for each beam momenta.
  With a help of theoretical models,
  an upper limit of the total decay width of $\Theta^{+}$ was estimated to be
  0.36 and 1.9~MeV for $J^{P}=1/2^{+}$ and $1/2^{-}$ states, respectively.
\end{abstract}
%\PACS{13.75.-n,13.75.Gx,14.20.Pt,25.80.Hp}

\section{Introduction}
Exotic hadrons are good tools to probe the internal degree of freedom of hadron
which is a composite particle made of quarks. Although exotic hadrons which consists of four or more quarks are allowed to exist in the framework of QCD,
they have never been observed in experimental searches.
Recently, the LHCb experiment reported observation of cryptoexotic pentaquark states on the $J/\psi p$ mass spectrum \cite{lhcb}. In addition,
a candidate for dibaryon system was found in measurements of $np$ scattering
by the WASA-at-COSY experiment \cite{wasacosy}.
In 2003, the first discovery of an explicitly exotic pentaquark was reported from the LEPS Collaboration \cite{leps}. In the same year, a tetraquark candidate was observed by Belle Collaboration \cite{bellex}.
The narrow widths of these states are of great interest, since they are possibly
a result of special internal structure.
The structure of such states are now still under debate. However,
it will deepen our understanding of ordinary hadrons to study the properties of exotic hadrons.

The first discovery of the pentaquark $\Theta^{+}$ was followed by many positive and negative experimental results. It should be noted that the existence is still not confirmed in most of repeated or dedicated experiments.
Under the circumstance, it is strongly desired to confirm the existence especially in a hadronic reaction in which a production cross section is expected to be higher than that in the photoproduction. Additionally, a high mass resolution is a key to realize high experimental sensitivity for a search for the narrow state.
The $\Theta^{+}$ was searched for in the $\pi^{-}p\rightarrow K^{-}X$ reaction as the first experiment at the J-PARC Hadron Facility.
It should be noted that
the pentaquark $\Theta^{+}$ was searched for in the same reaction in the KEK-PS E559 experiment \cite{e522}.
The upper limit of the production cross section was $2.9\mu$b, however the bump
structure was observed with a significance of $2.5\sigma$ which is not enough to claim the existence.
The J-PARC E19 experiment was performed in 2010 and 2012, with beam momenta of 1.92 and 2.01~GeV/$c$.
Accumulated number of beam pions are $7.8\times 10^{10}$ and $8.1\times 10^{10}$ in 2010 and 2012, respectively. The results of analyses are reported in literatures \cite{shirotori} and \cite{moritsu}. The summary of these results and future prospects are described in this paper.

\section{Experimental Apparatus}
The E19 experiment was performed at the K1.8 beamline at the J-PARC Hadron Facility.
The beamline provides separated secondary particles such as $\pi, K, \bar{p}$ up to 2~GeV/$c$ \cite{k18}.
The typical intensity of kaons is $10^{5}$~Hz, and that of $\pi$ is $10$M~Hz or higher.
At the very beginning of the beamline operation,
a duty factor of the primary beam was $10-20\%$, therefore an acceptable
averaged intensity of $\pi^{-}$ was $\sim1M$~Hz. Now the duty factor has been
improved to be $\sim50\%$, and the acceptable beam intensity is not limited
by the beam power but performances of experimental detectors.

\begin{figure}[htb]
  \centerline{%
    \includegraphics[width=5.5cm]{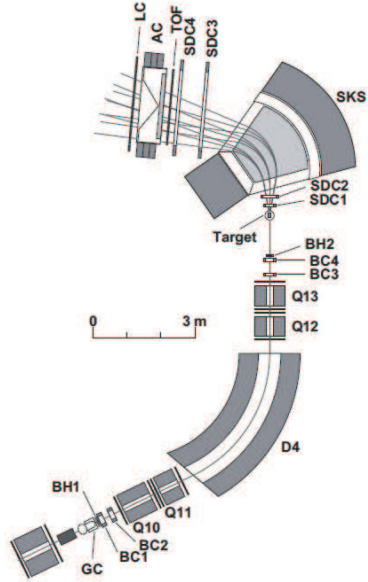}}
  \caption{Schematic view of K1.8 and SKS spectrometers.}
  \label{Fig:sks}
\end{figure}

The pentaquark $\Theta^{+}$ was searched for with the missing-mass technique
using the $\pi^{-}$ beam irradiated to a liquid hydrogen target.
The beam pions were analyzed with a K1.8 beamline spectrometer and scattered
kaons were detected with the Superconducting Kaon Spectrometer (SKS). Figure
\ref{Fig:sks} shows a schematic view of these two spectrometers.
The beam pions are identified with Time-of-Flight measurements and momentum
analyzed with a series of QQDQQ magnets together with MWPCs and DCs.
The $K^{-}$ are separated from $\pi^{-}$ with Time-of-Flight measurements.
The momentum of scattered $K^{-}$ was analyzed with DCs which were placed
at an entrance and exit of the SKS magnet.
The detail of the experimental apparatus is described elsewhere \cite{moritsu}.
%\begin{figure}[htb]
%  \begin{minipage}{\textwidth}
%    \begin{center}
%      \includegraphics[width=6.5cm,bb=0 0 568 872]{sks.jpg}
%    \end{center}
%    \caption{Schematic view of K1.8 and SKS spectrometers.}
%    \label{Fig:sks}
%  \end{minipage}
%  \begin{minipage}{0.5\textwidth}
%    \begin{center}
%    \end{center}
%  \end{minipage}
%\end{figure}

\begin{figure}[htb]
  \centerline{%
    \includegraphics[width=\textwidth]{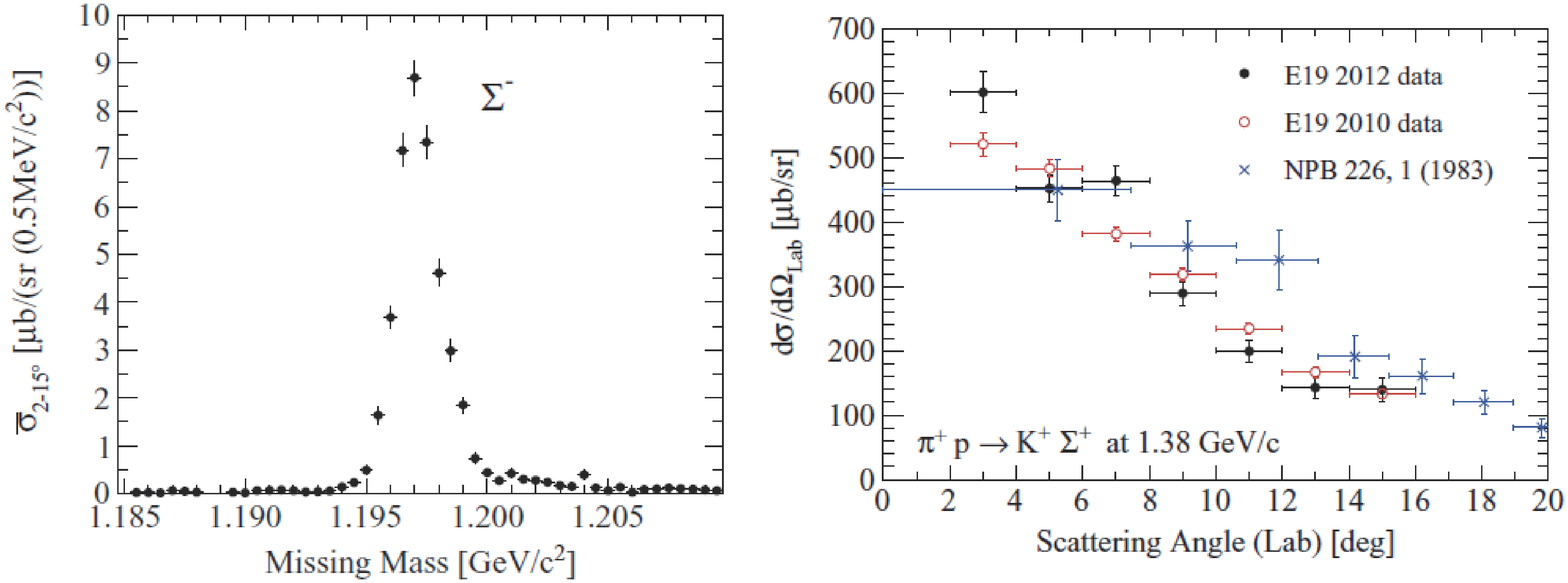}}
  \caption{(Left) Missing mass spectrum of $\pi^{-}p \rightarrow K^{+}X$ reaction.
    The clear peak corresponds to the ground state $\Sigma^{-}$.
    (Right) The production cross section of $\Sigma^{+}$
    measured in the 1.38~GeV/$c$ $\pi^{+}p\rightarrow K^{+}\Sigma^{+}$ reaction.
    Open and closed circles correspond to 2010 and 2012 data, respectively.
    The crosses show the result of the past experiment \cite{candlin}.
    The figures are taken from the literature \cite{moritsu}.
  }
    \label{Fig:sigma}
\end{figure}

The overall performance of spectrometers was examined using the
$\Sigma$ productions in the $\pi^{+}p\rightarrow K^{+}\Sigma^{+}$ and
$\pi^{-}p\rightarrow K^{+}\Sigma^{-}$ reactions.
Figure \ref{Fig:sigma} shows the missing-mass spectrum of the
$\pi^{-}p \rightarrow K^{+}X$ reaction and the production cross section
of $\Sigma^{+}$ measured in the $\pi^{+}p\rightarrow K^{+}\Sigma^{+}$ reaction.
A position of the observed peak is in a good agreement with the PDG value
taking into account the energy loss at the LH2 target.
The missing-mass resolution in the case of $\Theta^{+}$ production is
estimated to be 1.72 and $2.13$~MeV (FWHM) for the beam momenta of
1.92 and 2.01~GeV/$c$, respectively \cite{tomo,moritsu}.
The mass resolution of the former 2010 data was originally
reported as 1.4~MeV in \cite{shirotori}, however it turned out to be
1.72~MeV through an improved tracking algorithm
and systematic calibration method as described in \cite{tomo}.
The measured differential cross section of $\Sigma^{+}$ is consistent
with the past experimental result \cite{candlin}.
The more precise result has been obtained
in the forward region as shown in Fig. \ref{Fig:sigma} (right).

\section{Result and Discussion}
\begin{figure}[htb]
  \centerline{%
    \includegraphics[width=\textwidth]{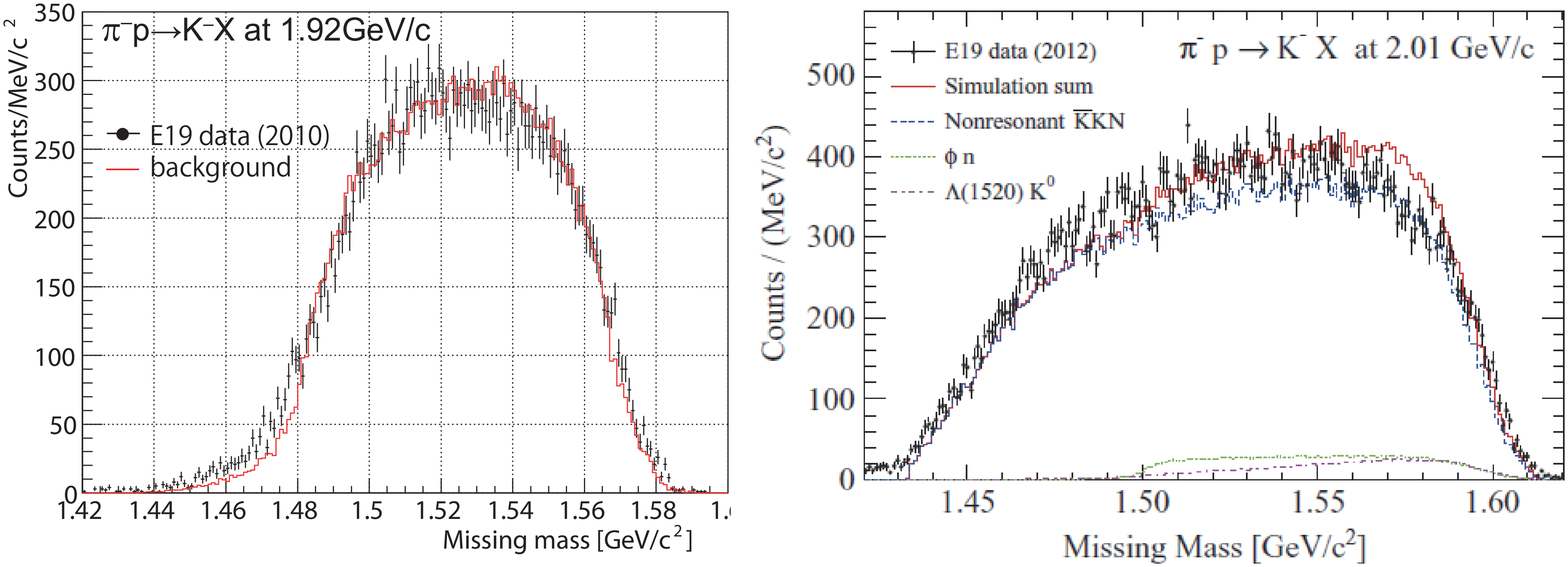}}
  \caption{Missing mass spectrum of $\pi^{-}p \rightarrow K^{-}X$ reaction
    with beam momenta of (left) 1.92 \cite{shirotori} and (right) 2.01~GeV/$c$ \cite{moritsu}.
    The simulated backgrounds are overplotted with histograms.
    In the right figure, background components,
    non-resonant $\bar{K}KN$ (dashed),
    $\phi$ meson (dotted) and $\Lambda(1520)$ productions (dot-dashed) are
    plotted. }
    \label{Fig:mass}
\end{figure}

\begin{figure}[htb]
  \centerline{%
    \includegraphics[width=\textwidth]{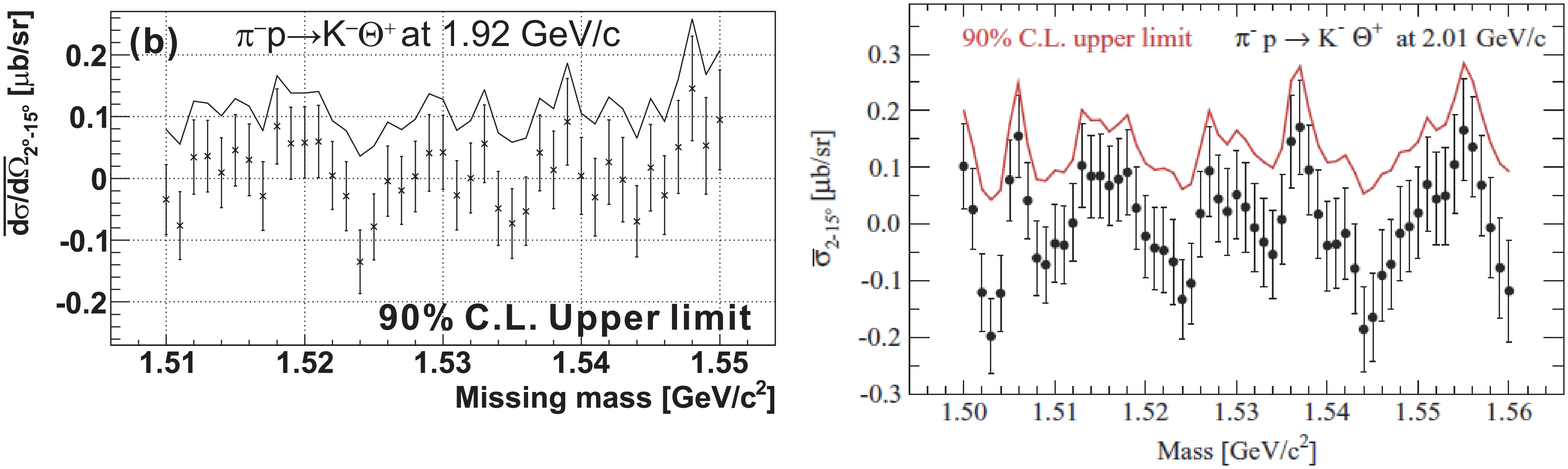}}
  \caption{The upper limits of production cross section
    at 90\% confidence level in the 1.92 (Left) and 2.01~GeV/$c$ (right)
   $\pi^{-}p\rightarrow K^{-}\Theta^{+}$ reactions \cite{shirotori,moritsu}.}
  \label{Fig:upperlimit}
\end{figure}

The missing-mass spectra of 1.92 and 2.01~GeV/$c$ $\pi^{-}p\rightarrow K^{-}X$ reactions are shown in Fig. \ref{Fig:mass}.
No significant peak was observed on the spectra.
Considerable background processes are $\phi$ and $\Lambda(1520)$ productions
together with a non-resonant $\bar{K}KN$ production. The cross sections
of these processes were measured in the past bubble chamber experiment \cite{dahl}.
A simulated background spectrum is overplotted on the plots showing
no peculiar structure in the mass region of $\Theta^{+}$.
It should be noted that the cross section of non-resonant $\bar{K}KN$
is scaled to present data otherwise the amplitude of background
is underestimated by a few times.
The 90\% confidence level upper limits of the production cross section of
$\pi^{-}p\rightarrow K^{-}\Theta^{+}$ reaction are plotted in Fig. \ref{Fig:upperlimit}.
The upper limits are estimated to be 0.28$\mu$b/sr in the laboratory frame in the mass region of $1.51-1.55$~GeV/$c^{2}$ for both 1.92 and 2.01~GeV/$c$
$\pi^{-}p\rightarrow K^{-}\Theta^{+}$ reactions \cite{tomo,moritsu}.
The upper limit obtained in the present experiment is smaller than that of the
previous KEK experiment \cite{e522} by an order of magnitude.
The measured upper limit of the cross section can be translated into
the upper limit of decay width of $\Theta^{+}$ with a help of
model dependent theoretical calculations. The result of some
calculations with different coupling schemes and treatments of
form factors is well summarized in \cite{hyodo2}.
With 2010 and 2012 data combined, the upper limit of decay width
was obtained to be 0.36~MeV and 1.9~MeV for different spin states of
$J^{P} = 1/2^{+}$ and $1/2^{-}$, respectively,
around the $\Theta^{+}$ mass region of $1.53-1.54$~GeV/$c^{2}$.
In the case of $J^{P} = 1/2^{+}$, our upper limit is just same as
the decay width of 0.34~MeV which was derived from the measurement of $\Theta^{+}$ in $K^{+}A$ reactions reported by DIANA Collaboration \cite{diana3}.
A comparison with the LEPS results is minutely discussed in the literature \cite{tomo}. Currently there seems no theoretical framework to explain the finite
cross section reported by LEPS Collaboration within our upper limit of decay width.

\section{Future Prospects}
In the $\pi^{-}p\rightarrow K^{-}X$ reaction, the $\Theta^{+}$ might
possibly be generated through an intermediate state like a cryptoexotic $N^{*}$
as suggested by CLAS Collaboration in the $\gamma p\rightarrow \pi^{+}K^{-}K^{+}n$ reaction \cite{clasp}.
The mass of the state is around 2.4~GeV/$c^{2}$, which is unfortunately above
the maximum $\sqrt{s}$ of the $\pi^{-}p$ reaction at the K1.8 beamline.
The possibility of production through the exotic $N^{*}$ can be examined with
a higher secondary beamline which is available in near future at J-PARC.

At present, there is a few survived positive results.
Recently new analysis of a data set for 2006$-$2007 was presented
by the LEPS Collaboration
reporting that
the strong peak of pentaquark $\Theta^{+}$ was not reproduced in the same
manner originally applied to the old data \cite{kato}.
However, it was found that the peak structure seems to be clearly enhanced
when applying a new method to reject events from spectator protons.
They continue data taking so far and a new result is eagerly awaited.

At the J-PARC Hadron Facility, no experiment dedicated for
the search for the $\Theta^{+}$ is planned in the next couple of years.
However, several experimental studies concerning exotic hadrons will be
performed in the near future.
An experiment has been just started to investigate the spectral information
of $\Lambda(1405)$. Currently a new beamline which delivers higher momentum
particles is under construction.
It is named as the high-momentum beamline (high-p) providing
primary protons up to 30~GeV/$c$.
It is also designed as the secondary beamline up to 20~GeV/$c$
by inserting a production target.
A spectroscopy of $\Xi$ and charmed baryons has recently been proposed
at the new beamline,
since the higher beam momentum enable us to study these unexploited part of
baryons.

\section{Summary}
The pentaquark $\Theta^{+}$ was searched for with the missing mass technique
in the 1.92 and 2.01~GeV/$c$ $\pi^{-}p\rightarrow K^{-}X$ reactions
at the K1.8 beamline of the J-PARC Hadron Facility.
The $\Sigma^{\pm}$ productions were studied in the pion-induced reactions
to evaluate the spectrometer performances.
No significant peak structure was observed on the missing-mass spectra of
$\pi^{-}p\rightarrow K^{-}X$ reaction.
The 90\% confidence level upper limit of the production cross section of
$\pi^{-}p\rightarrow K^{-}\Theta^{+}$ is estimated to be 0.28$\mu$b/sr
in the laboratory frame. The obtained upper limit of the decay width
is 0.36 and 1.9~MeV for $J^{P} = 1/2^{+}$ and $1/2^{-}$, respectively.

\bibliographystyle{plain}
\bibliography{naruki}

\end{document}